\documentclass[intlimits,twoside,a4paper]{article}

\usepackage[cp1251]{inputenc}
\usepackage{multirow}
\usepackage[eqsecnum]{cmpj3}

\usepackage{bm}

\newcommand{\ps}{\,\textrm{ps}}
\newcommand{\fs}{\,\textrm{fs}}

\issue{2023}{26}{3}{33605}
\doinumber{10.5488/CMP.26.33605}

\title[Curcumin solutions in water-DMSO]
{Aspects of the microscopic structure of curcumin solutions with water-dimethylsulfoxide solvent. 
Molecular dynamics computer simulation study}
\author[T. Patsahan, O. Pizio]{T. Patsahan\orcid{0000-0002-7870-2219}\refaddr{label1}\refaddr{label2},
 O. Pizio\orcid{0000-0001-8333-4652}\refaddr{label3}
\thanks{Corresponding author: \email{oapizio@gmail.com}.}}
\addresses{
    \addr{label1} Institute for Condensed Matter Physics of the National Academy of Sciences of Ukraine,\\
    1 Svientsitskii St., 79011 Lviv, Ukraine
   \addr{label2} Institute of Applied Mathematics and Fundamental Sciences, Lviv Polytechnic National University, 12~S.~Bandera Str., UA-79013 Lviv, Ukraine
    \addr{label3} Instituto de Qu\'{i}mica, Universidad Nacional Aut\'{o}noma de M\'{e}xico, 
   Circuito Exterior, 04510, Cd. Mx., M\'{e}xico}

\date{Received April 4, 2023, in final form May 23, 2023}

\begin{document}

\maketitle

%%%%%%%%%%%%%%%%%
\begin{abstract}
 We explore some aspects of the microscopic structure of curcumin solutions with 
water-dimethylsulfoxide solvent of variable composition. Molecular dynamics computer 
simulations at isobaric-isothermal conditions are used for this purpose.
The model consists of the OPLS-UA type model for the enol conformer of curcumin
 (J. Mol. Liq., \textbf{223}, 707, 2016), the OPLS model for the dimethylsulfoxide (DMSO)
and the SPC/E water model.
Radial distributions for the centers of mass of curcumin molecules are evaluated 
and the corresponding running coordination numbers are analyzed. 
The disaggregation of curcumin clusters upon increasing the DMSO content in  water-DMSO  
solvent is elucidated.
Changes of the distribution of water and DMSO species
 around curcumin molecules are investigated. A qualitative comparison of our findings with the results
 of other authors is performed. A possibility to relate predictions of the model
 with the experimental observations in terms of the so-called critical water
aggregation percentage is discussed.  
\printkeywords
%
%
%\pacs 02.70.Ns,61.20.Ja,82.30.Rs,87.15.hp
\end{abstract}

\section{Introduction}

This manuscript has been prepared as a tribute to Prof. Taras Bryk, distinguished Ukrainian
scientist in the field of statistical physics, computer simulations 
of water, aqueous solutions and interfaces, on behalf of his 60th birthday. 
Dr.~T.~Bryk has made several important contributions along different lines of research in
the theory of liquids. One part of his past and recent studies,
see e.g.,~\cite{bryk1,bryk2,bryk3,bryk4,bryk5},
is methodologically related to the objectives of the present work,
specifically to molecular dynamics computer simulations of water and complex solutions.

In this short communication, we are interested in liquid solutions containing a
small amount of curcumin solutes.
In general terms, it is an ample scientific problem 
of interest for computer modelling and applications that require systematic investigations.
We would like to report here only some very fresh results, yielding certain
basic insights within this kind of research. There remains much room for further studies.

Curcumin is of much interest due to pharmacological aspects of its application,
see e.g.~\cite{ghosh1,kumar1,luthra1,mehanny1,ngo1}. However, this bioactive 
hydrophobic polyphenol
belongs to the group of poorly water soluble drugs. Thus, its delivery to the targets
for therapeutic purposes is problematic \cite{ucisik}.  
In some cases, solvent blends have been used to develop liquid formulations~\cite{lv,cui} 
with enhanced solubility of curcumin solutes and improved delivery properties.
In contrast to water solvent, curcumin is quite soluble in organic solvents such as
the DMSO, ethanol and methanol, for example.
Therefore, for exploratory purposes, in this study we use
the solvent consisting of two species, namely of water and DMSO
with variable composition. DMSO is the aprotic polar solvent of much interest for
biological applications on its own~\cite{quinn}. 

From the theoretical perspective, the system in question is extremely complex. 
In order to uncover the fundamental issues and provide hints for experimental studies,
the quantum chemistry methods and molecular dynamics computer simulations 
have been applied for curcumin solutions. Due to various simplifying assumptions
intrinsic within theoretical modelling and focus on solely specific issues of
the problem, a comparison of the results with 
experimental predictions remains elusive~\cite{wright1}. 

Quite recent studies of curcumin solutions by using computer simulation techniques
\cite{ngo2,Varghese-2009,Wallace-2013,Samanta-2013,Hazra-2014,Sreenivasan-2014,Yadav-2014,Parameswari-2015,Priyadarsini-2009} in the majority of cases restrict to pure water as solvent. 
Much less
attention has been paid to the performance of curcumin in non-aqueous solvents.
Moreover, the simulations in many cases restricted to the models
of a single curcumin molecule in the ``sea'' of solvent species. Thus, a competition
between the effects of solute-solute and solute-solvent interactions has been left 
out of attention.
Commonly, the modelling of the curcumin molecule force field comes out from the quantum chemical (QC)
calculations within the B3LYP method using different versions of Gaussian software~\cite{Samanta-2013,Hazra-2014,bonab,pereira}.

Recently, Ilnytskyi et al. developed a simpler, OPLS-united atom (UA) model
for the curcumin molecule~\cite{Ilny-2016} using the OPLS data basis~\cite{OPLS-1996}.
The model for a single molecule was tested in vacuum and in water medium 
using classical molecular dynamics simulations~\cite{Ilny-2016}.
Next, for the sake of comparisons, the model was tested in methanol and in DMSO 
apart of  water, by using  molecular dynamics simulations~\cite{Pat-2017}.
A qualitative agreement for the specific trends of behaviour of the systems in question
with the modelling of other authors was obtained.
Further, in the most recent work from this laboratory~\cite{Pat-2021},
we focused in systems with a larger number of
curcumin molecules ($N_{\text{cur}}=2$, $4$, $8$, $12$, $16$ and $20$)
in water, in close similarity to the studies of other authors~\cite{Hazra-2014,bonab,pereira}. 
At this stage, our predictions for the microscopic
structure yielding self-assembly of curcumin solutes into clusters are
discussed versus the results of ~\cite{Hazra-2014} concerned with the
similar computational setup.

In order to specify the principal objective of the present work, we would like to 
make a few comments pertinent to the expected behaviour of the systems under study.
%The curcumin molecule is hydrophobic, in part due to the presence of phenyl rings. 
As documented in \cite{Hazra-2014} from the analysis provided in \cite{hamaguchi}, 
the hydrophobic nature of the phenyl
rings of curcumin along with hydrogen bond formation ability of the side groups 
is proposed to be the reasons for the aggregation in water. 
It is difficult to discern the observed trends as separate contributions from
the direct interaction between phenyl rings, determined by the model, 
and/or from the cooperative effect 
of aqueous medium. Definitely, water solvent promotes
hydrophobic association of curcumin molecules and leads to the formation of
clusters. 
Intuitively, to get a more homogeneous solution at the microscopic scale, or say
to mitigate the undesired clustering, one should 
substitute a certain amount of water molecules by organic co-solvent species.
We used here the DMSO molecules with this aim.
The DMSO co-solvent is miscible with water. Amphiphilic character of the 
DMSO molecule is due to two methyl groups in its chemical structure. Affinity of these groups
to the elements of the curcumin molecule  contributes
to the solubility of curcumin in pure DMSO~\cite{lv,cui}. However, these groups
``dislike'' water species. 
On the other hand, the oxygen atom of DMSO molecule can form hydrogen bonds with either 
of water hydrogens. 
Thus, the amphiphilicity of DMSO results in the hydrophobic effect in water-DMSO mixture 
at a low DMSO content in water~\cite{aguilar}.
This effect is more pronounced in water-DMSO mixture compared to mixtures of  water 
with simple alcohols~\cite{cruz}.
In summary, in qualitative terms, one would expect that clusters of curcumin molecules will 
self-destroy or disaggregate upon adding the DMSO species to water. 
Apparently, the principal mechanism of declusterization 
would come from the permeation of DMSO molecules into loosely packed curcumin clusters 
and of solvation of curcumin clusters and monomers by DMSO, that both lead to 
less and less attractive effective interaction, or equivalently of the
less attractive potential of the mean force  between the curcumin molecules. 
However, this effective interaction depends on the distance and on orientation of curcumin 
molecules. We intend to analyze these trends in terms of distance dependent 
descriptors for the phenyl groups of curcumin molecules and of some atom-atom 
pair distribution functions.

This kind of behaviour
was elucidated very recently in~\cite{pereira} for curcumin dissolved in
water-ethanol mixtures and characterized in terms of useful, 
although rather crude property, referred to as the critical water aggregation 
percentage, i.e., the limit percentage in which curcumin monomers are predominant
rather than their self-aggregated state. This percentage was 
abbreviated as CWAP. 
The authors performed concise  spectroscopic analysis of curcumin solutions
with water-ethanol solvent using chemometric software
to capture the CWAP and discuss various related issues.
These observations guide us to explore the existence of CWAP and its
evaluation for curcumin in water-DMSO solvent by using analysis of the microscopic structure 
upon changing the solvent composition. Unfortunately, we have not found experimental data
in this respect for the solutions that involve the DMSO species.

\section{Model and simulation details}

Molecular dynamics computer simulations of curcumin molecules in water-DMSO
solvent with variable composition  have been performed
in the isothermal-isobaric (NPT) ensemble at a temperature of $298.15$~K and at $1$~bar.
The GROMACS simulation software \cite{GROMACS} version 5.1.4 was used.

\begin{figure}[!ht]
\centering
\includegraphics[width=0.5\textwidth,angle=0,clip=true]{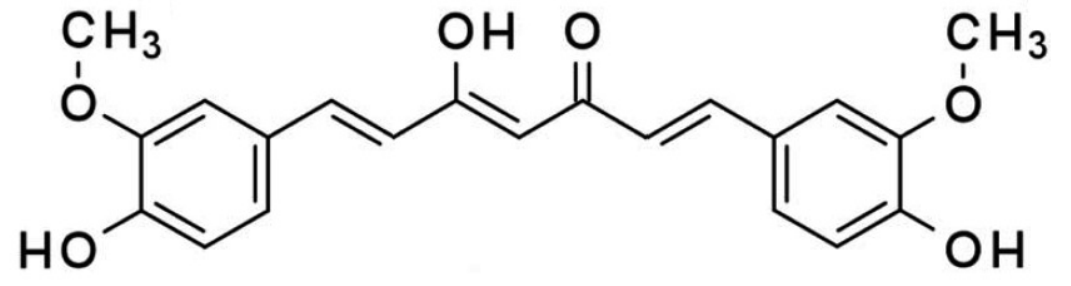}
%\vspace{3em}
\caption{Chemical structure of the enol form of curcumin molecule.}
%figure 1
\label{fig_enol}
\end{figure}

\begin{figure}[!ht]
\centering
\includegraphics[width=0.5\textwidth,angle=0,clip=true]{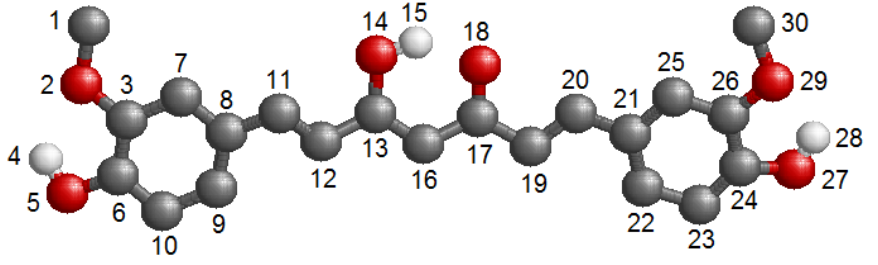}
%\vspace{3em}
\caption{(Colour online) Scheme for the united-atom
curcumin model with sites numbering. Carbon groups are shown as dark gray spheres,
oxygens --- as red spheres, hydrogens --- as small light-gray spheres.}
\label{fig_curc}
\end{figure}

The model for curcumin molecule was constructed in the previous studies 
from this laboratory by using 
the OPLS-united atom (UA) force field~\cite{Ilny-2016}.
Water and DMSO are described in the framework of SPC-E model~\cite{SPCE}
and the OPLS-UA (as in~\cite{Pat-2017}), respectively.
This combination of the model for three constituents of the solution 
had been already  tested  by considering a single curcumin molecule in water, 
methanol and DMSO~\cite{Pat-2017}.
Similarly to our previous works, we restrict attention to the enol tautomer of curcumin.
This structure is dominant in solids and in various solvents~\cite{slabber,kolev}.
The chemical structure of the molecule is given in figure~\ref{fig_enol}.
The ``ball-and-stick'' schematic representation of the molecule
had been  already presented  in \cite{Ilny-2016,Pat-2017,Pat-2021}. Still, 
for the sake of convenience of the reader, we show
it here again in figure~\ref{fig_curc}.

%The ``ball-and-stick'' schematic representation of the molecule
%has been presented already in \cite{Ilny-2016,Pat-2017,Pat-2021}. Still, we show
%it here again in Fig.~\ref{fig_curc}, for the sake of convenience of the reader.

All the parameters of the force field for curcumin are given in the supporting information
file to~\cite{Ilny-2016}. All bonds in the curcumin model as well as for DMSO, 
are considered as rigid and accounted for via constraints,
using the LINCS algorithm implemented in GROMACS. 

Technical details of the simulation procedure are the same as in \cite{Pat-2017,Pat-2021}.
%and are repeated here for convenience of the reader.
The geometric combination rules were applied to determine the parameters for the cross interactions
(rule~3 of GROMACS software). To evaluate Coulomb interactions contributions, the particle
mesh Ewald procedure was used (fourth-order spline interpolation and grid spacing 
for the fast Fourier transform equal to $0.12$).
The cut-off distance both for Coulomb and Lennard-Jones interactions was chosen equal to $1.1$ nm.
The van der Waals tail correction terms to the energy and pressure were taken into account.

For each system, a periodic cubic simulation box was set up with $N_{\text{tot}}=4000$ molecules,
$N_{\text{tot}}= N_{\text{cur}}+N_{\text{wat}}+N_{\text{DMSO}}$. Principal results coming from
our calculations concern  the systems containing twelve curcumin molecules,
$N_{\text{cur}}$ = $12$. In the series of runs  we explored the effects
of the composition changes of the solvent. Only the final result for the CWAP
is given for the twice bigger number of curcumin molecules, $N_{\text{cur}}$ = $24$, as well.

\begin{figure}[!ht]
\centering
\includegraphics[height=4.8cm,clip]{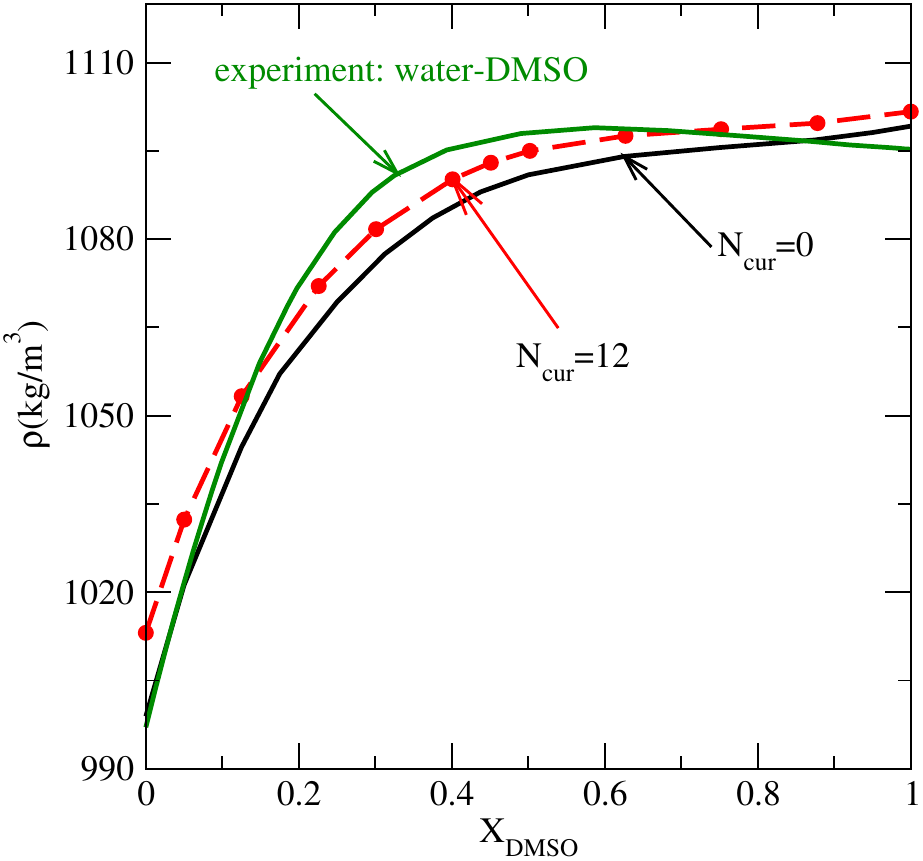}
\caption{(Colour online) The dependence of density of water-DMSO
mixture on composition from experimental data~\cite{egorov} and
from simulations using the OPLS-UA model for DMSO and SPC/E water
(green and black lines, respectively). Composition dependence of
%(independent on the
%presence or absence of clusters in the system with $N_{\text{cur}}$ = $12$)
density  of the systems with $N_{\text{cur}}$ = $12$ and $N_{\text{DMSO}} + N_{\text{wat}}$=3988
solvent molecules from simulations (red line with circles). 
In all cases, simulations are performed at the temperature 298.15~K and the pressure 1~bar.
%In all cases, $298.15$~K and at $1$~bar. 
The circles correspond to all 
simulated systems of this work with different composition of the DMSO-water solvent. 
}
\label{fig_dens}
\end{figure}

The initial configuration of particles was prepared by placing first
the $N_{\text{cur}}$ molecules randomly in the simulation box. Next, $N_{\text{DMSO}}$ and $N_{\text{wat}}$ molecules
were added into the box.
Each of the systems  underwent the energy minimization to remove
a possible overlap of atoms in the starting configuration. This was done by applying the steepest
descent algorithm. After the minimization procedure, we performed equilibration of each system with
the time duration $10$~ps at $298.15$~K and $1$~bar using a small time step $0.1$~fs.
The Berendsen thermostat with the time constant of $0.1$~ps and isotropic
Berendsen barostat with the time constant of $2$~ps were applied.
The compressibility parameter was taken equal to $5.25\cdot 10^{-5}$ bar$^{-1}$
in all simulations. 

After equilibration, the production runs were performed. We used the V-rescale thermostat
with the time constant of $0.5\ps$ and Parrinello-Rahman barostat with the time
constant of $2$~ps, implemented in the
GROMACS software, to perform production runs. The time step in production runs 
was chosen to be $1\fs$.
The results were collected from the production run of $150$~ns.
For the sake of convenience of the reader, in figure~\ref{fig_dens}, we show the composition dependence
of density for water-DMSO mixture coming from the models of water and of DMSO used 
in the present study.
The experimental curve is shown as well. Finally, the nominal density
(independent of the cluster existence) of the systems
with 12 curcumin molecules and 3988 solvent molecules on composition, resulting
from the NPT simulations is given a the dashed curve. 
A set of circles corresponds to all the systems with different composition under study.

We would like to finish this section by a set of comments concerning the choice
of the solvent model. For purposes of the present study, it would be desirable to apply 
the most precise model for water-DMSO mixture that captures the hydrophobic effect
and accurately describes the deviation of various mixing properties from ideality in the entire
interval of compositions. In spite of the efforts of many groups, an overall agreement
between the computer simulation results and experimental data is not entirely satisfactory
at the level of non-polarizable water-DMSO models. Detailed analyses of the state of the art
was discussed in our recent contributions~\cite{aguilar,gujt}. 
The DMSO and curcumin modelling both are from the OPLS-UA database. 
It is known that the DMSO OPLS-UA model combined with
the SPC/E water is reasonably good for the description of energetic
aspects of mixing of solvent species between themselves, cf. figure~3 of~\cite{gujt}. 
However, this combination of models is not very accurate 
for the geometric aspects of mixing, say for the excess mixing volume,
and does not permit to capture the hydrophobic effect in terms of e.g., apparent
molar volumes of species~\cite{aguilar}. 
In summary, it is desirable to employ a more sophisticated
model for the mixed solvent, possibly including the TIP4P-type water
model and a more sophisticated DMSO description, in future studies of the problem.

\newpage
\section{Results and discussion}

\subsection{Composition trends of dis-assembly of curcumin particles in water-DMSO solvent. Overall insights}

\begin{figure}[!ht]
 \centering
\includegraphics[height=6.0cm,clip]{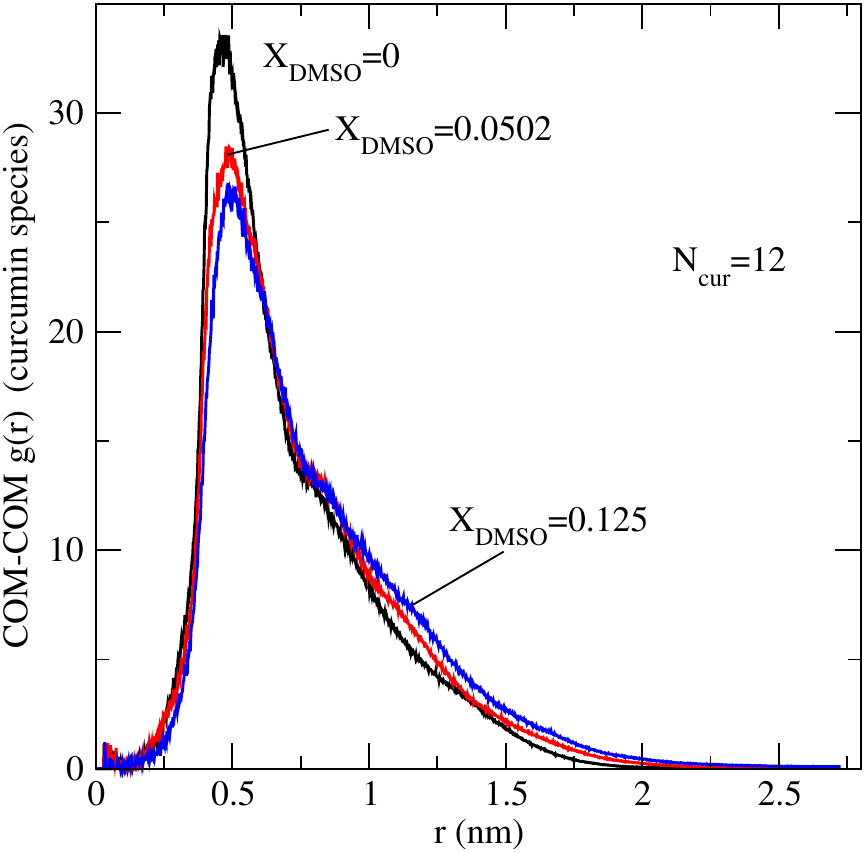}
\includegraphics[height=6.0cm,clip]{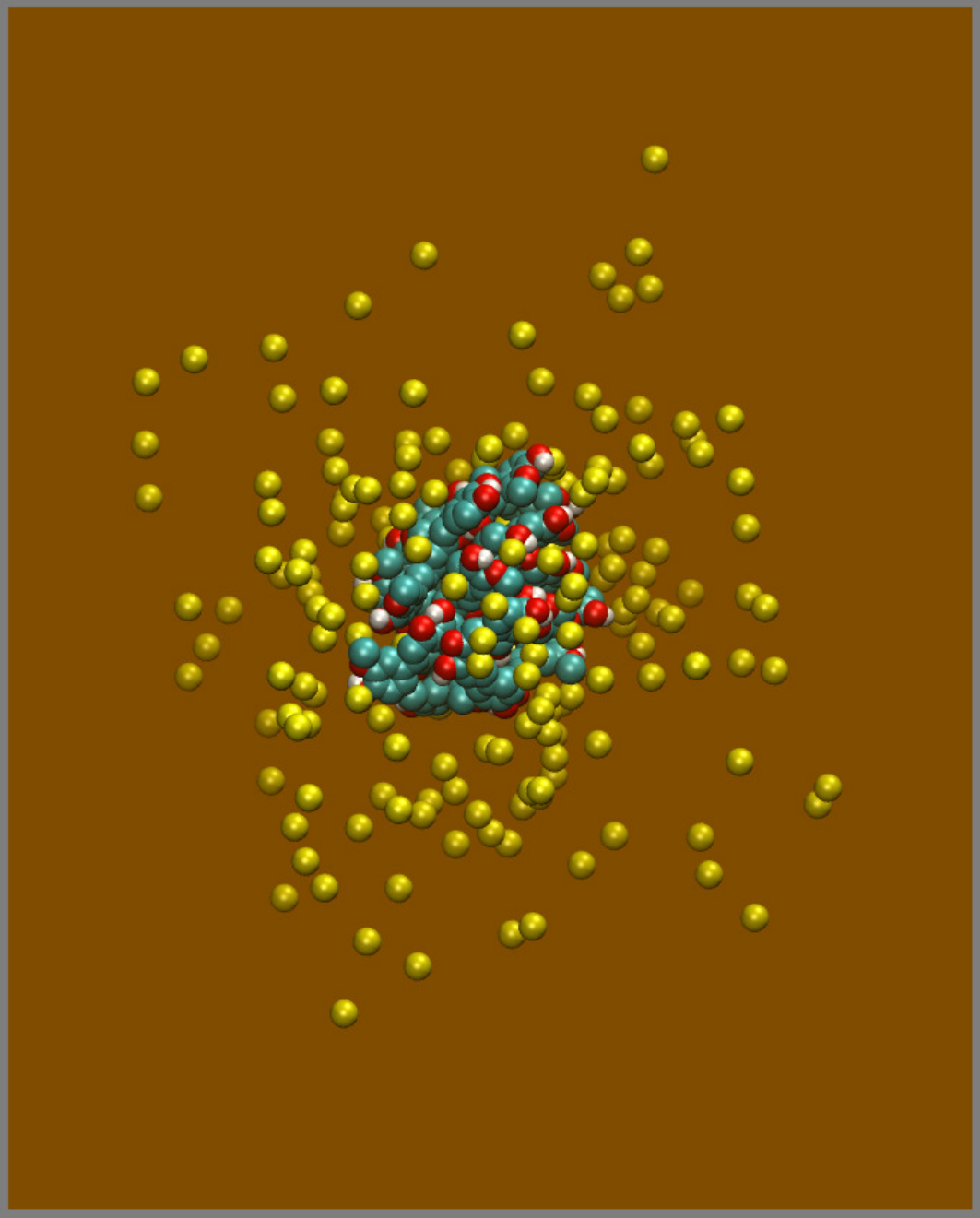}
\caption{(Colour online) Left-hand panel: Changes of the
COM-COM radial distribution of curcumin molecules in water-DMSO solvent 
upon changing the composition at a low DMSO content. Right-hand panel:
Snapshot of the cluster of curcumin molecules at $X_{\text{DMSO}} = 0.0502$.
The DMSO sulfur atoms are plotted as yellow spheres.}
\label{fig4_rdf_panel_a}
\end{figure}

We begin with the description of the radial distribution 
of the center of mass (COM) of curcumin species in the 
solvent of variable composition.
From the left-hand panel of figure~\ref{fig4_rdf_panel_a}, 
we learn that  $12$ curcumin molecules self-assembly into a cluster or 
clusters, in close similarity to the pure water solvent case in~\cite{Pat-2021},
if the solvent contains a small fraction of DMSO molecules, because the radial distributions
shown in this panel decay to zero at a certain interparticle separation. 

We apply the geometric definition for a cluster. Namely, taking into account the behaviour of 
the radial distribution of curcumin molecules in pure water explored 
by us recently \cite{Pat-2021}, we choose here the ``cut-off'' 
COM-COM distance $R_{\text{c}}=0.75$~nm, 
as a criterion for the cluster definition.
Hence, all the curcumin molecules satisfying this geometric criterion 
during the production run are counted as belonging to a cluster.

The radial distributions for the systems in question
are characterized by a single, not very sharp,
maximum, at around $0.45$~nm and small shoulder around $0.8$~nm. In pure water solvent,
with the same number of solute molecules, although this shoulder
is not observed. The height of the maximum decreases, if the number of the DMSO molecules
in the solvent increases or equivalently the number of water molecules decreases. The width of the
upper part of the radial distribution apparently does not change upon increasing the DMSO amount in
the solvent. However, the shoulder develops and the width of the radial distribution in its lower part
increases when $X_{\text{DMSO}}$ increases from $0$ to $0.125$  ($X_{\text{DMSO}} = N_{\text{DMSO}}/N_{\text{tot}}$).
One may interpret this behaviour
in terms of the changes of the internal structure of a cluster. It attains a certain internal structure
and becomes slightly more ``diffuse'' because the DMSO molecules permeate the cluster.  

Moreover, it can be seen from the typical snapshot in the right-hand panel of figure~\ref{fig4_rdf_panel_a}, 
for $X_{\text{DMSO}}= 0.0502$, that
certain fraction of the DMSO molecules fills the periphery of the cluster or in other words
solvates it. Other DMSO molecules are quite uniformly distributed within water.  
This and the following  plots for snapshots were prepared by using the VMD software~\cite{VMD}.

\begin{figure}[!ht]
 \centering
\includegraphics[height=6.0cm,clip]{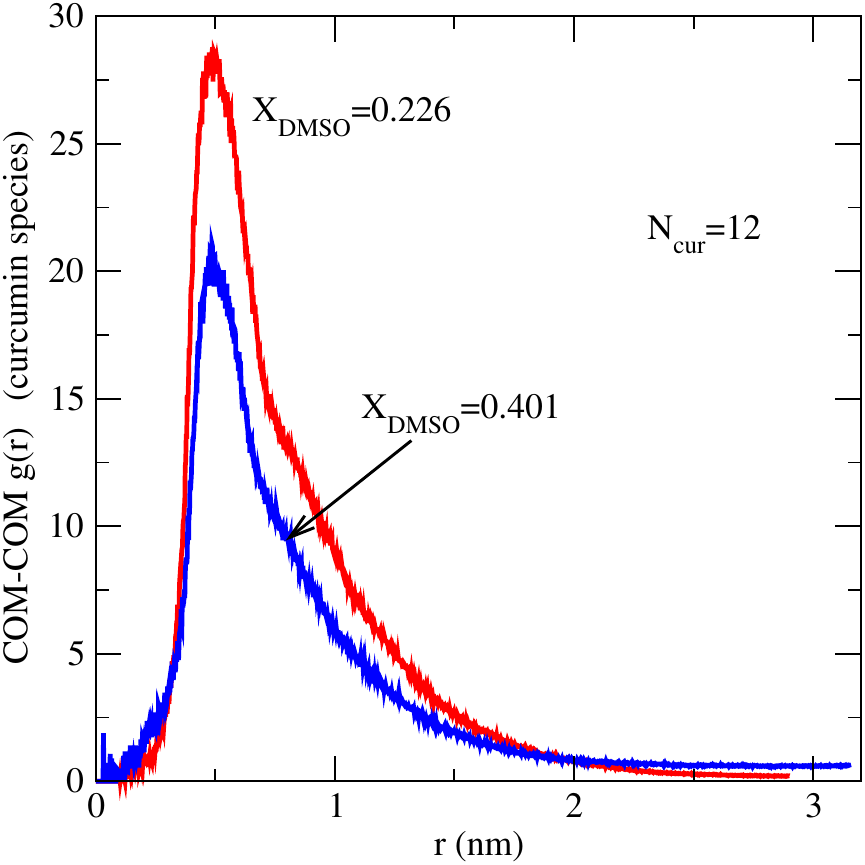}\\
\includegraphics[height=6.0cm,clip]{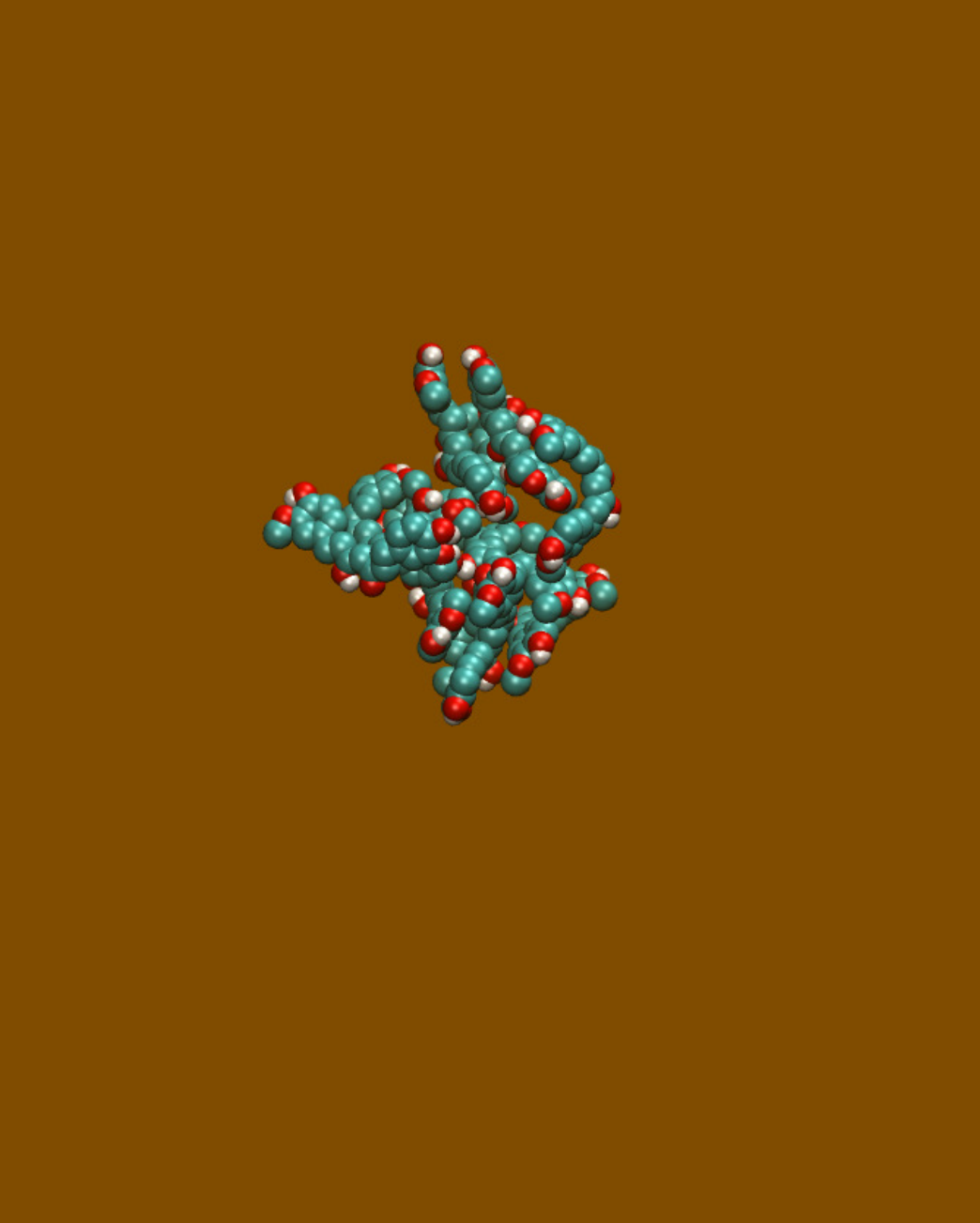}
\includegraphics[height=6.0cm,clip]{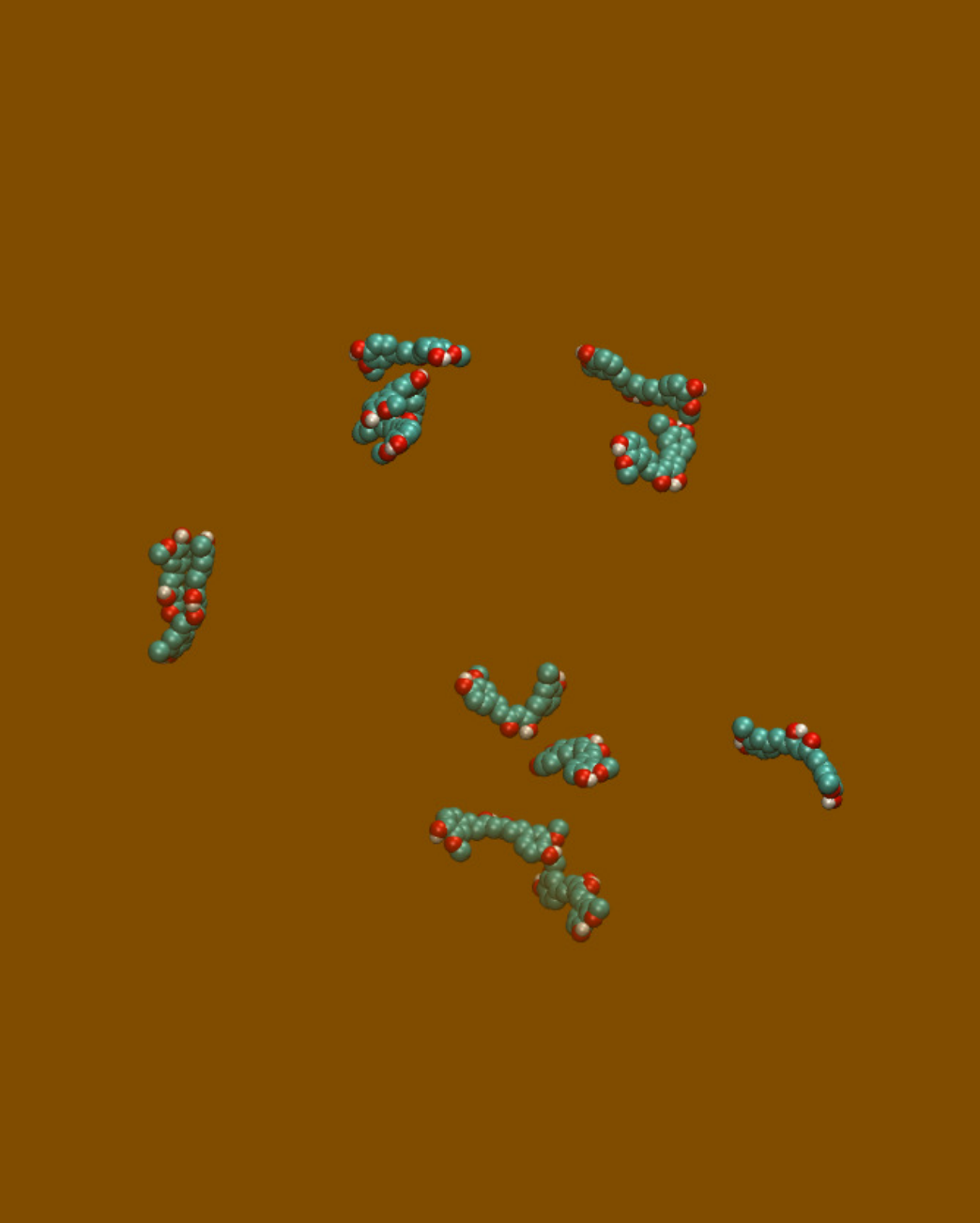}
\caption{(Colour online) Top panel: Changes of the
COM-COM radial distribution of curcumin molecules in water-DMSO solvent upon changing
composition at intermediate  DMSO content. Bottom panels: Snapshot of the curcumin
cluster at $X_{\text{DMSO}} = 0.226$ and $X_{\text{DMSO}} = 0.401$, respectively. 
A few monomers are detached from the cluster at $X_{\text{DMSO}} = 0.226$,
cf. the histogram distribution below in figure~10a.
In the latter case, $X_{\text{DMSO}} = 0.401$, monomers and dimers are observed, 
cf. the histogram distribution in figure~10b.
}
\label{fig5}
\end{figure}

Upon further increase of the DMSO amount in the solvent, figure~\ref{fig5}, we observe that
at a lower concentration, $X_{\text{DMSO}} = 0.226$, the cluster is well defined, cf. the radial 
distribution function in panel~a and the snapshot in panel~b. A few monomers may
detach from the cluster, as it follows from the snapshot. Still, the radial distribution
tends to zero at distances larger than 2.5~nm. 
At a higher DMSO concentration, $X_{\text{DMSO}} = 0.401$, the structure of curcumin species 
is very different in comparison with the previous case. The radial distribution changes the 
asymptotic behaviour (top panel of figure~\ref{fig5}), it tends to a constant, less than unity, 
describing a more homogeneous distribution. Still, as it follows from the snapshots in figure~\ref{fig5}, 
the curcumins exhibit microheterogeneity. Some of the curcumin monomers show trends for
pairing into dimers.

\begin{figure}[!ht]
 \centering
\includegraphics[height=5.5cm,clip]{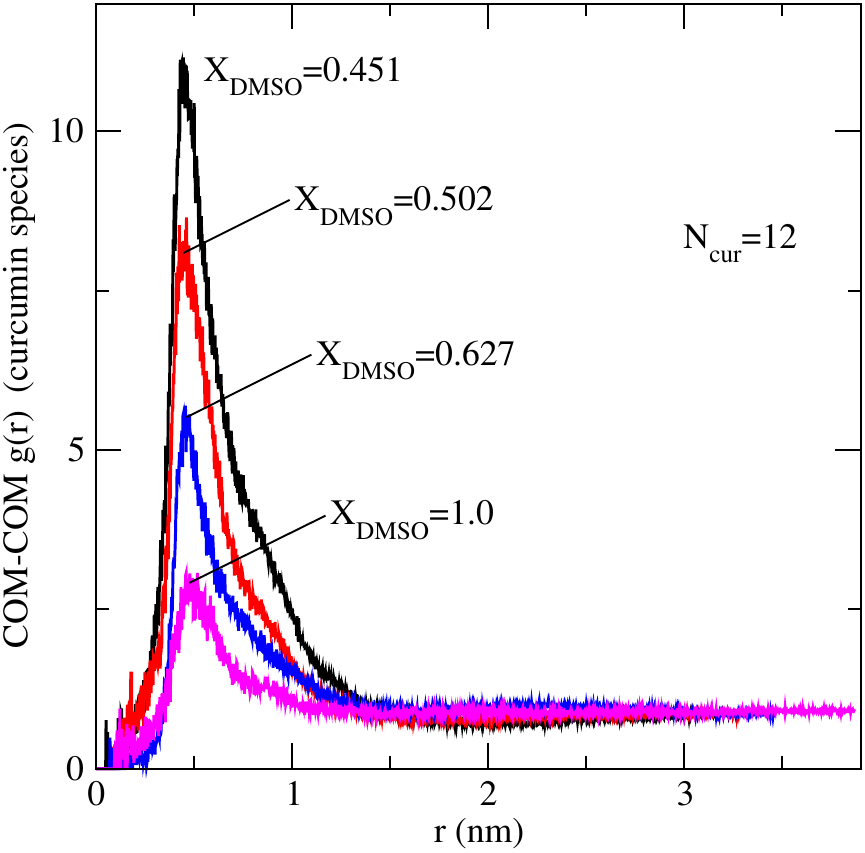}
\caption{(Colour online) Changes of the
COM-COM radial distribution of curcumin molecules in water-DMSO solvent 
upon changing the composition in the interval of a high DMSO content.}
\label{fig6}
\end{figure}

If the solvent is characterized by even higher concentration of DMSO species, 
the first and only maximum of the radial distribution is observed, figure~\ref{fig6}.
The height of the maximum  decreases with increasing $X_{\text{DMSO}}$. 
Moreover, the maximum becomes narrower. 
Apparently, the shape of the COM-COM radial distribution  
without minimum after the first maximum and any oscillations is due to a very low
curcumin concentration. The distribution of curcumin species is uniform after the first maximum.

\begin{figure}[!ht]
 \centering
\includegraphics[height=5.5cm,clip]{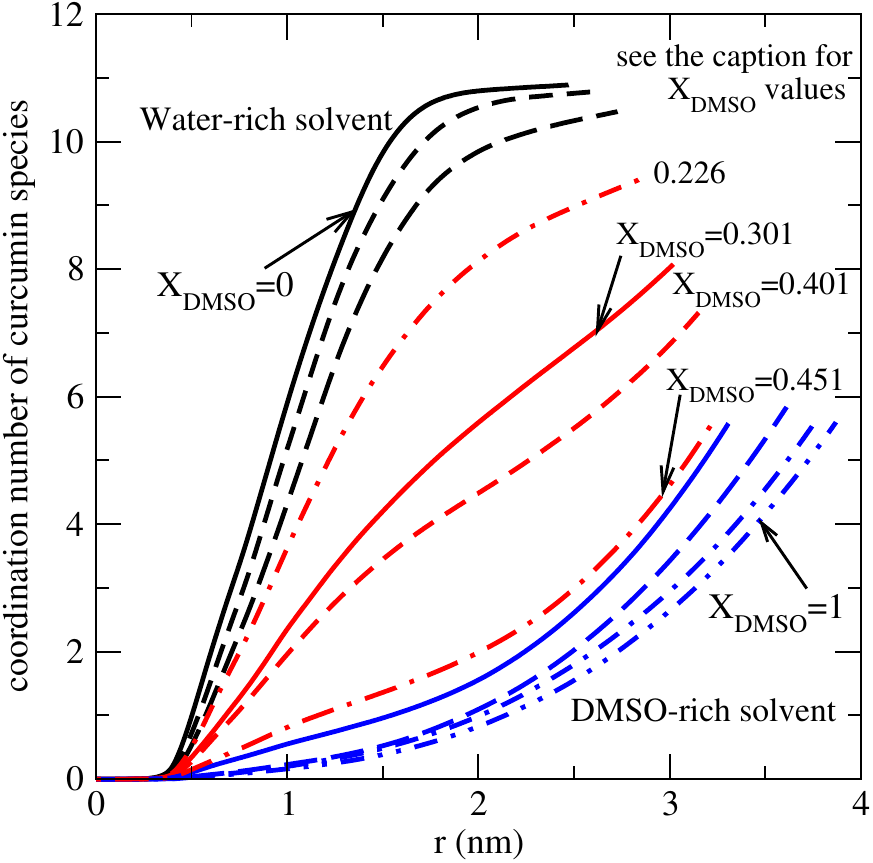}
\caption{(Colour online) Illustration of changes of the coordination number of
curcumin species in water-DMSO solvent with variable
composition. The curves from the top to bottom correspond to
$X_{\text{DMSO}}$ = $0$, $0.0502$, $0.125$, $0.226$, $0.301$, $0.401$, $0.451$, $0.502$,
$0.752$, $0.878$ and $X_{\text{DMSO}}=1$. Some of the curves possess labels for better
visualization.}
\label{fig7}
\end{figure}

Trends of possible aggregation of curcumin molecules into a cluster in the entire
composition interval can be
conveniently followed by the evolution of the running coordination number for the COM
of curcumin molecules. It is defined as follows,
\begin{equation} 
  n_{\text{cur-cur}}(r) = 4\piup(N_{\text{cur}}/L^3) \int_0 ^r  g(R)  R^2 \rd R,
\end{equation}
where $L$ is the length of the edge of the simulation box. The corresponding curves are given in
figure~\ref{fig7}. Saturation of the curves to the value, $N_{\text{cur}}-1$, 
at a low $X_{\text{DMSO}}$ indicates aggregation of curcumin 
species into a cluster.
On the other hand, at high values of $X_{\text{DMSO}}$, the coordination number monotonously grows. 
In terms of this indicator, a change of the shape of the curves occurs in the interval
between  $X_{\text{DMSO}} \approx 0.3$ and $X_{\text{DMSO}} \approx 0.4$. This interval corresponds to
the maximum deviation of the pure solvent properties from ideality, 
see e.g.,~\cite{aguilar,gujt}. 
The solvent composition with $X_{\text{DMSO}} \approx 0.3$ corresponds to 
one DMSO per two H$_2$O molecules approximately. 

\begin{figure}[!ht]
 \centering
\includegraphics[height=5.5cm,clip]{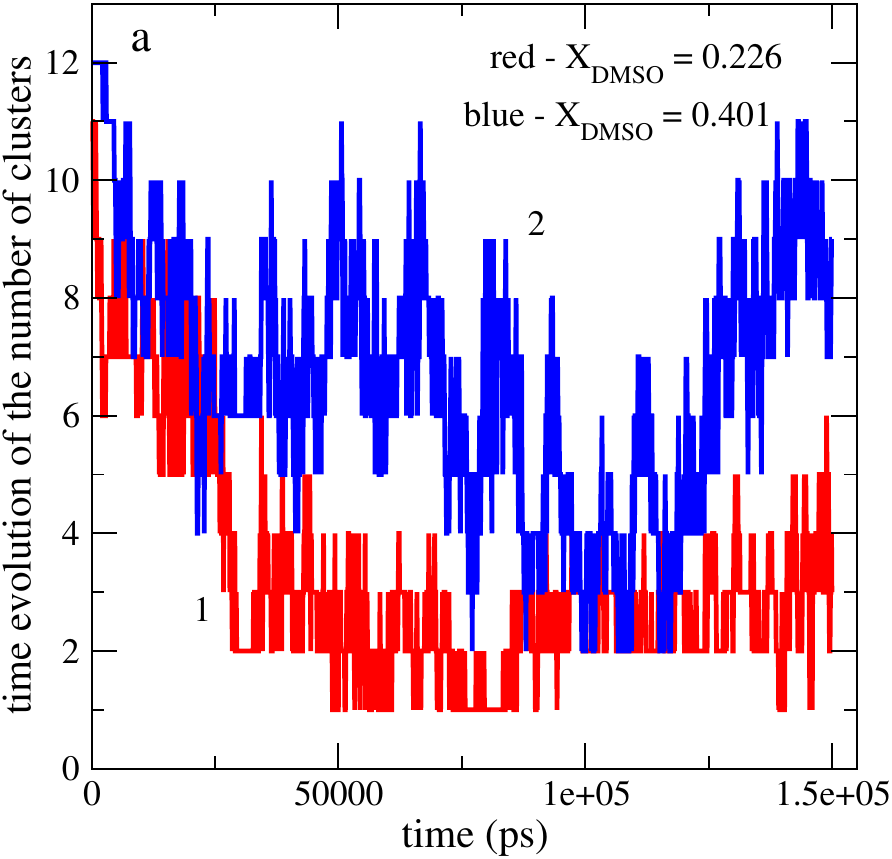}
\includegraphics[height=5.5cm,clip]{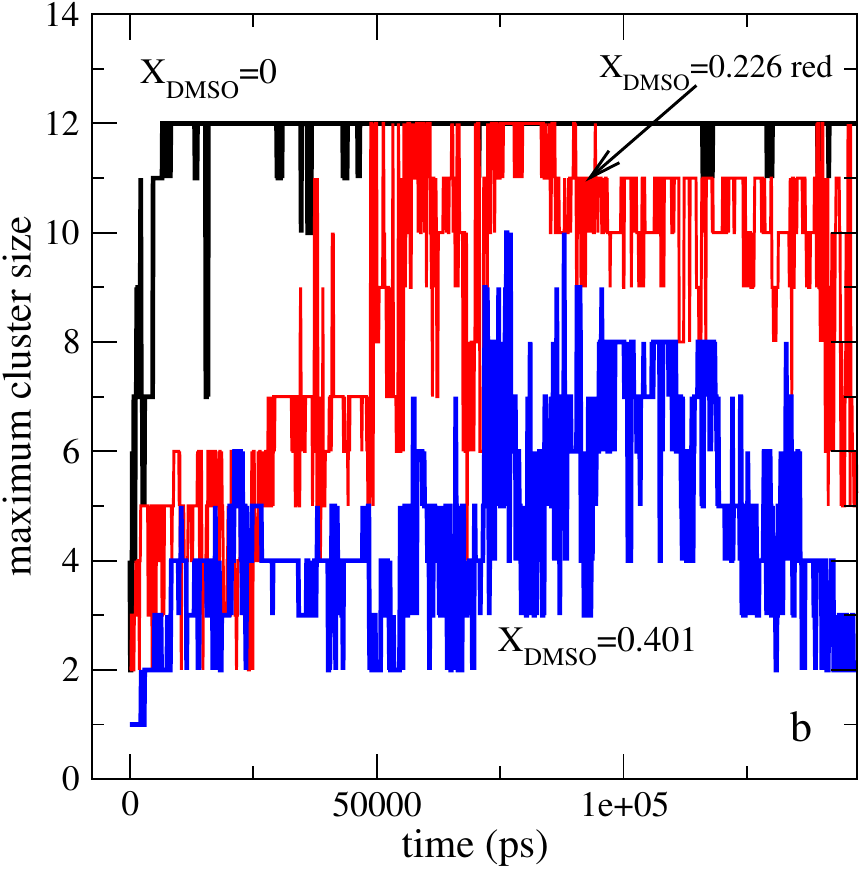}
\caption{(Colour online) Panel~a: Time evolution of the number of clusters 
of curcumin molecules ($N_{\text{cur}}$ = $12$)
at different concentration of DMSO species in the solvent.
Panel~b: Time evolutions of the maximum cluster size of curcumin species.
The values of $X_{\text{DMSO}}$ are given in both panels of the figure.
}
\label{fig8}
\end{figure}

\begin{figure}[!ht]
 \centering
\includegraphics[height=5.5cm,clip]{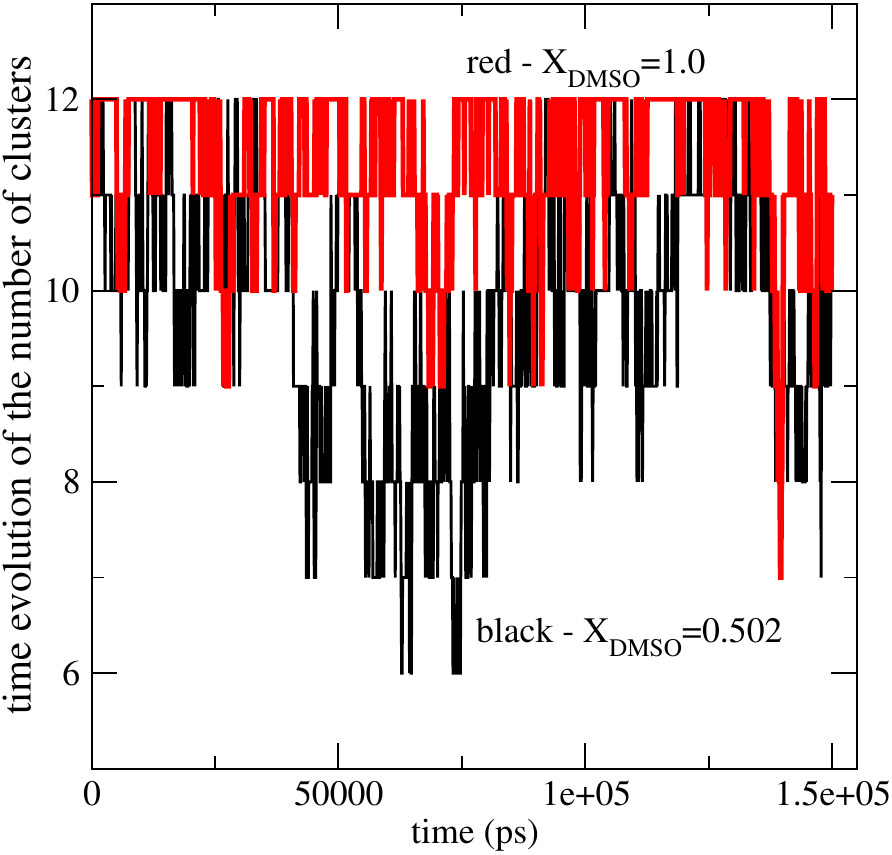}
\caption{(Colout online) Time evolution of the number of  clusters
of curcumin molecules ($N_{\text{cur}}$ = $12$)
at a high fraction of DMSO species in the solvent (black) and in pure 
DMSO (red).}
\label{fig9}
\end{figure}

The coordination number of curcumin species provides a quite general but
not profound insight into their aggregation. A  more detailed knowledge
can be extracted from the trajectories by exploring temporal changes
of the number of clusters during the system evolution. The time evolution 
of the number of clusters in the composition interval of out interest 
is described in panel~a of figure~\ref{fig8}. We observe that at a lower
content of DMSO in the solvent, $X_{\text{DMSO}} = 0.226$, the system evolves such
that during the final stage of the trajectory the number of clusters 
predominantly fluctuates in the interval between two and four. Hence, the
trimers and larger clusters exist under a chosen DMSO concentration. 
By contrast, at a higher DMSO fraction, $X_{\text{DMSO}} = 0.401$, the number
of clusters on the final stage of the production run is larger. At this
DMSO concentration, the dimers and possibly trimers prevail. Additional
insights follow from the evolution of the maximum cluster size, panel~b of figure~\ref{fig8}.
In pure water, the formation of a single cluster with twelve curcumin 
molecules is statistically the most frequent event. In the presence of  
DMSO species in the solvent, the maximum cluster of curcumin particles 
may consist of several entities, from $5$ to $11$, if $X_{\text{DMSO}} = 0.226$.
At a higher DMSO content, $X_{\text{DMSO}} = 0.401$, the maximum cluster size evidences 
the presence of dimers, trimers and tetramers at the end of the trajectory.
At even higher fraction of the DMSO species and in pure DMSO solvent, figure~\ref{fig9},
the number of clusters is higher than in the previous cases in figure~\ref{fig8},
showing a rather uniform distribution of curcumin molecules in the system,
although with appreciable fluctuations. Probably, longer runs are necessary 
to get better statistical trends.

\begin{figure}[!ht]
 \centering
\includegraphics[height=5.5cm,clip]{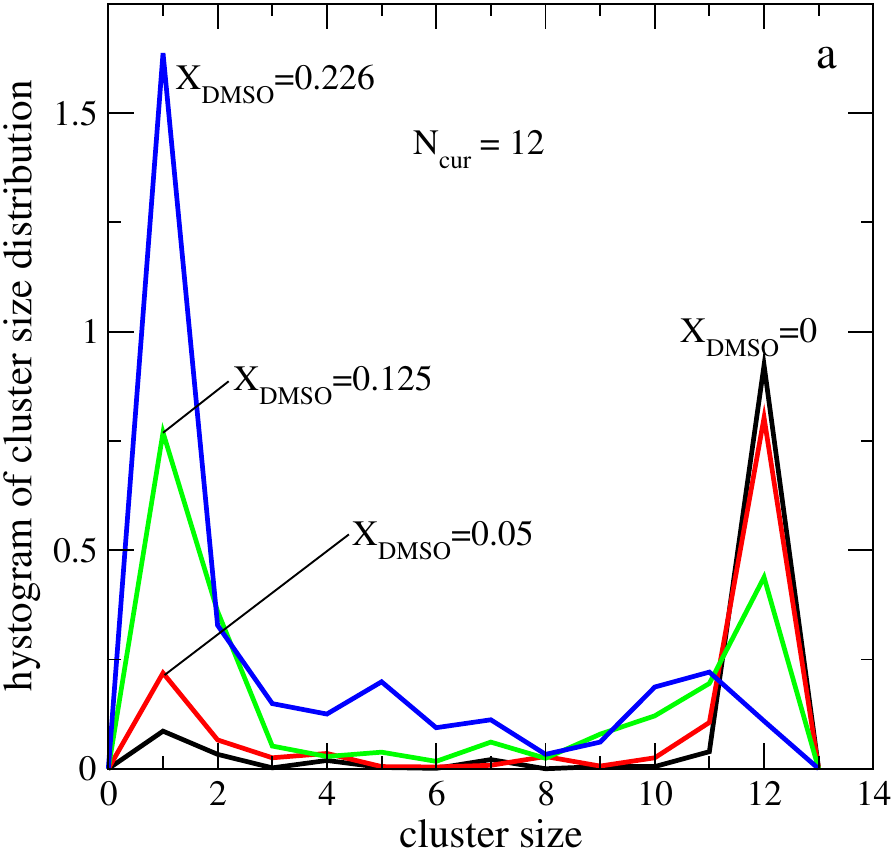}
\includegraphics[height=5.5cm,clip]{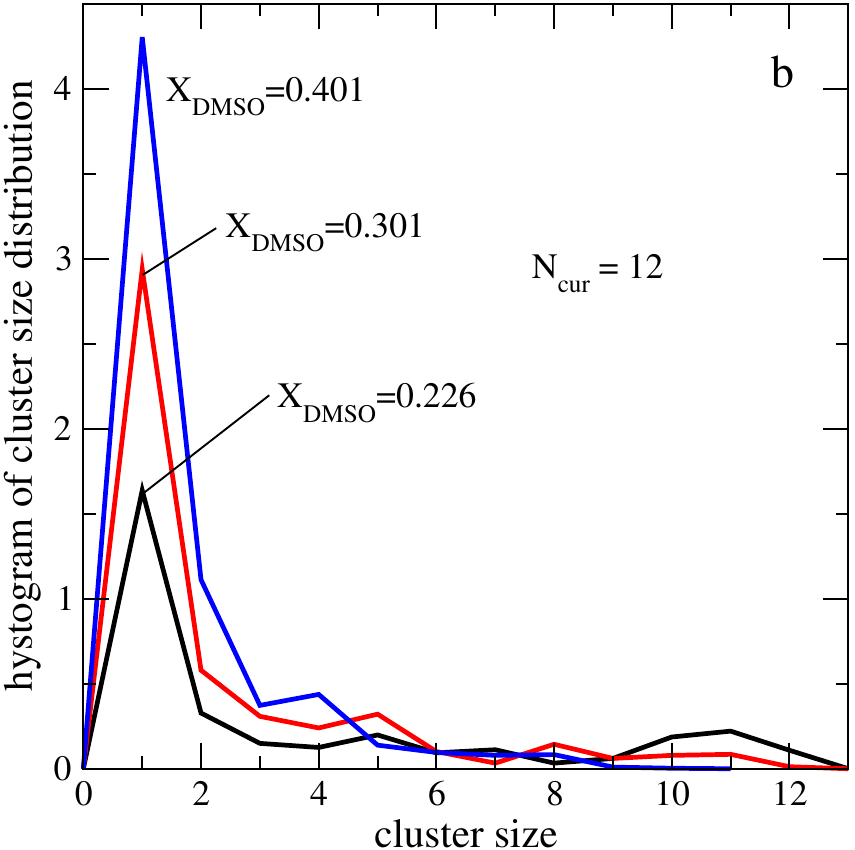}
\caption{(Colour online) Histograms of the cluster size distribution
at different concentration of DMSO species in the solvent.
The values of $X_{\text{DMSO}}$ are given in both panels of the figure.
}
\label{fig10}
\end{figure}

In order to describe the size distribution of clusters formed upon changes of
the solvent composition, we collected statistics of cluster size 
within the production run and constructed a set of histograms following from
the entire trajectories. The corresponding curves are given in
two panels of figure~\ref{fig10}. The following trends are worth mentioning.
A large cluster formation is dominant only at a very low DMSO content 
in the solvent. Already at $X_{\text{DMSO}} = 0.125$, the frequency of the  appearance
of monomers is slightly higher than the probability of the formation of a large curcumin cluster.
At a higher fraction of DMSO species in the solvent, $X_{\text{DMSO}} = 0.226$,
the frequency of monomers becomes higher than the appearance of
clusters of several curcumin molecules. Still, these clusters are observed
(panel~a of figure~\ref{fig10}). At even higher values for the DMSO fraction (panel~b of 
figure~\ref{fig10}), monomers prevail whereas the formation of clusters becomes much
less frequent. Actually, only dimers are frequent in comparison to larger $n$-mers.
Disaggregation of curcumin clusters upon increasing DMSO fraction in the solvent
results in structural changes in the solute particles distribution and
within the curcumin molecules surrounding, say in solute-solvent distribution.
Some observations concerned with these issues are illustrated in the following
figures.

\subsection{Trends of microscopic structure upon disaggregation of curcumin clusters}

\begin{figure}[!ht]
 \centering
\includegraphics[height=5.5cm,clip]{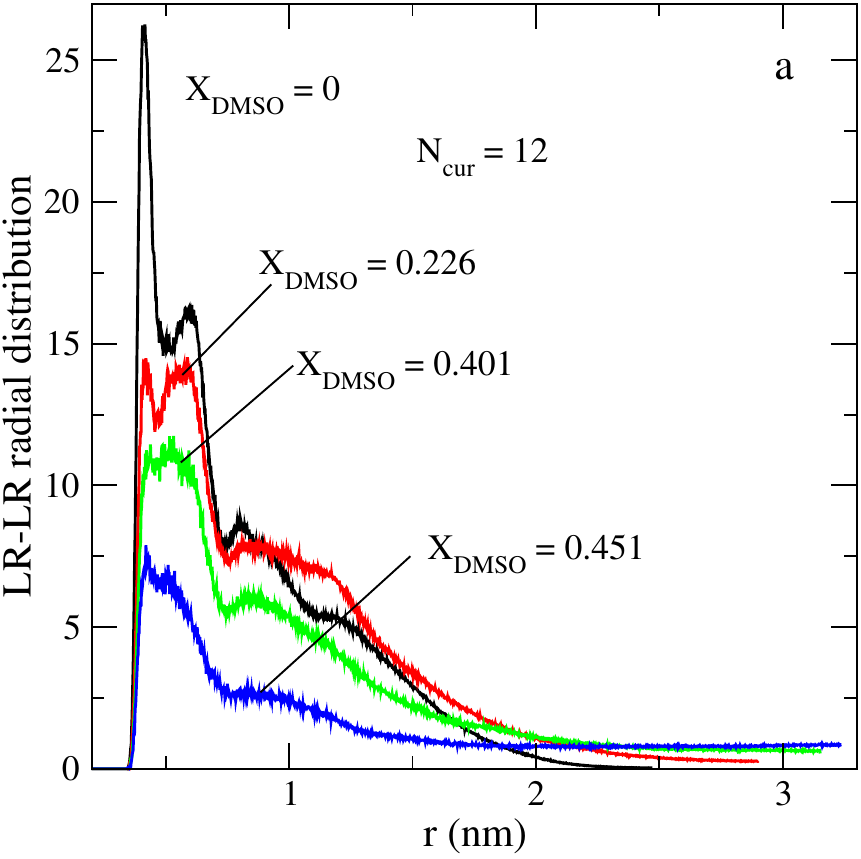}
\includegraphics[height=5.5cm,clip]{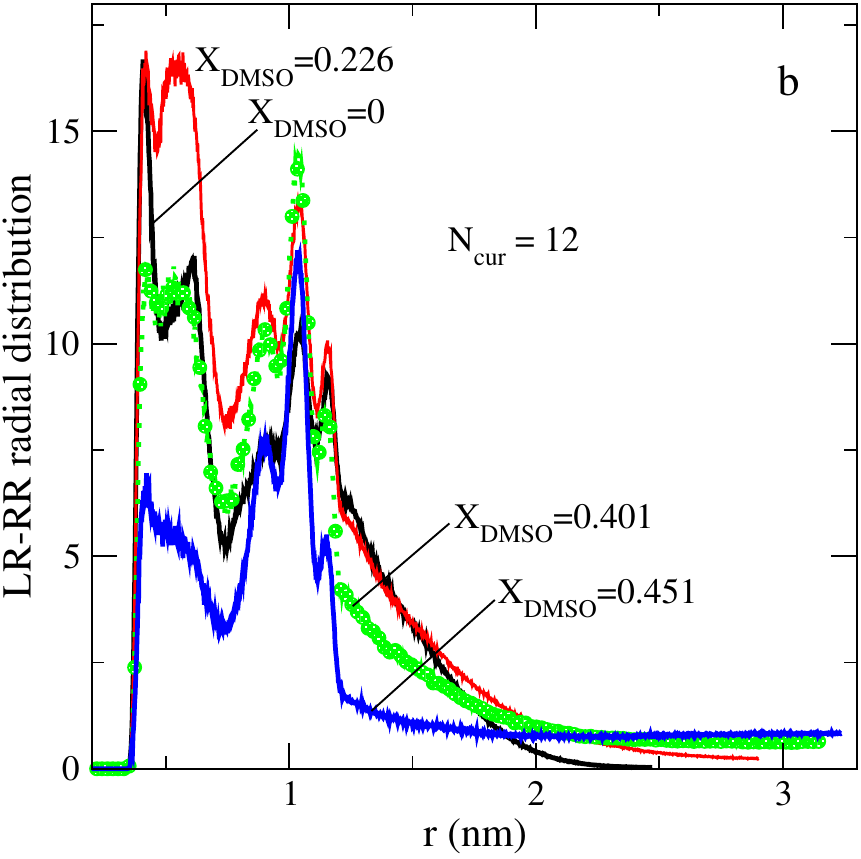}
\caption{(Colour online) Radial distribution of curcumin molecule phenyl rings
($N_{\text{cur}}$ = $12$)
at different concentration of DMSO species in the solvent.
The values of $X_{\text{DMSO}}$ are given in both panels of the figure.
}
\label{fig11}
\end{figure}

Changes of the internal structure of clusters
during their disaggregation in water-DMSO
solvent are illustrated in figure~\ref{fig11}. Namely, the radial distributions 
of phenyl rings (left-left, LR-LR, and left-right, LR-RR) 
of a curcumin molecule w.r.t. each other are shown.

The parallel orientation of the left-hand rings predominates at
low fraction of DMSO species in water, i.e., when a rather big
curcumin cluster is formed (panel~a of figure~\ref{fig11}). 
Substitution of a certain amount of water by the DMSO species results 
in a quite drastic decay of the magnitude and extent of this type of correlations. Thus, the 
orientation order of this type is  strongly diminished, when the DMSO fraction in
the solvent blend increases. The distribution  becomes rather uniform even at not
very big distances between curcumin molecules.

On the other hand, the anti-parallel orientation observed in water
and in water-DMSO solvent behaves differently (panel~b of figure~\ref{fig11}).
Interpretation of the curves in this panel is not straightforward, because
they involve an intramolecular contribution. We  already know that the intramolecular
LR-RR structure of a single curcumin molecule in water and in DMSO is
characterized by the distances 0.89~nm, 1.05~nm and 1.16~nm, cf. figure~5 in
each of the \cite{Ilny-2016,Pat-2017}, respectively. It is shown
that the characteristic distance 1.05~nm is the most probable.
All three characteristic distances are seen in all the curves given in panel~b of figure~\ref{fig11}.
  
However, additional features are as follows.
The closest ``contact''  value apparently does not change much,
if the DMSO content increases from $X_{\text{DMSO}} = 0$ to $X_{\text{DMSO}} = 0.226$. In the
latter case, it is even enhanced compared to $X_{\text{DMSO}} = 0$. 
Additional maximum at a low LR-RR separation may indicate the shift of a ring with
respect to the reference ring leading to parallel-displaced configuration
in the composition interval around $X_{\text{DMSO}} = 0.226$. At higher
DMSO fractions, this behaviour practically disappears.
In general terms, an
increase of the DMSO fraction results in a lower probability to find
anti-parallel orientation of phenyl rings. The features corresponding to
the intramolecular structure in fact are much less sensitive to the solvent
composition changes.
In addition, 
%the well pronounced maximum 
%of this distribution function  at $r \approx 1.1$nm preserves in a ample 
%interval of DMSO fractions, from $X_{\text{DMSO}} = 0$  to $X_{\text{DMSO}} = 0.451$. 
diminishing values of the distribution function in the interval of distances between 0.4~nm and 1~nm
may indicate the presence of the solvent separated phenyl rings as a result 
of disaggregation of clusters of curcumin species. 
At $X_{\text{DMSO}} > 0.4$, the distribution of rings at separation above 1.2~nm
becomes almost uniform.

\begin{figure}[!ht]
 \centering
\includegraphics[height=5.5cm,clip]{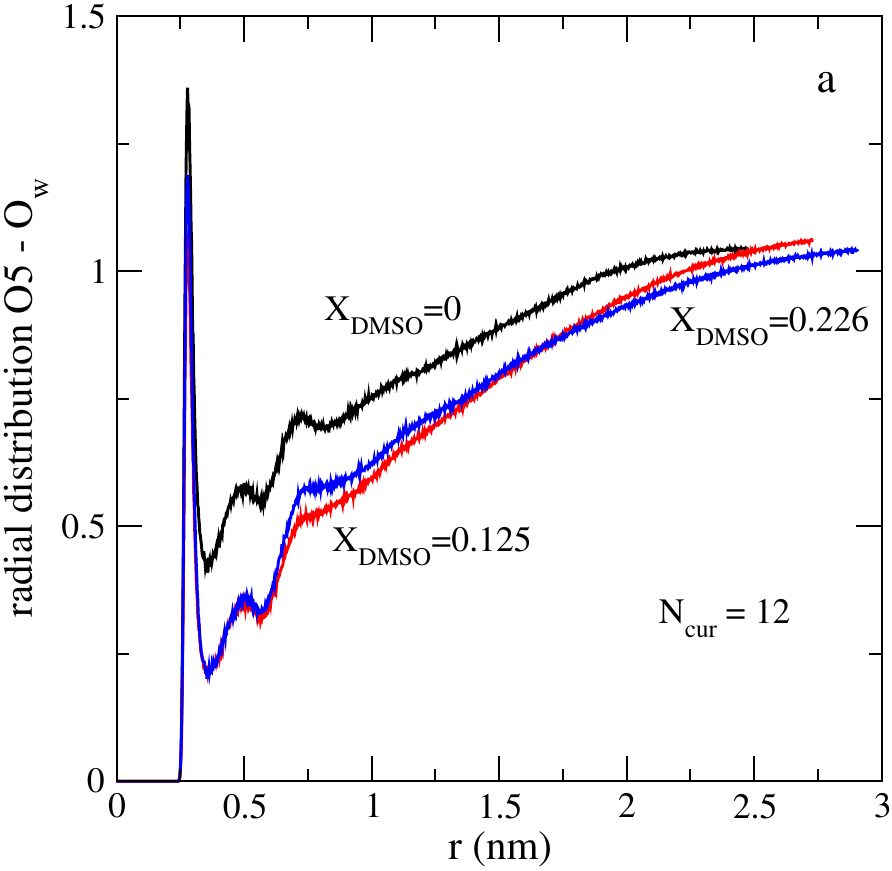}
\includegraphics[height=5.5cm,clip]{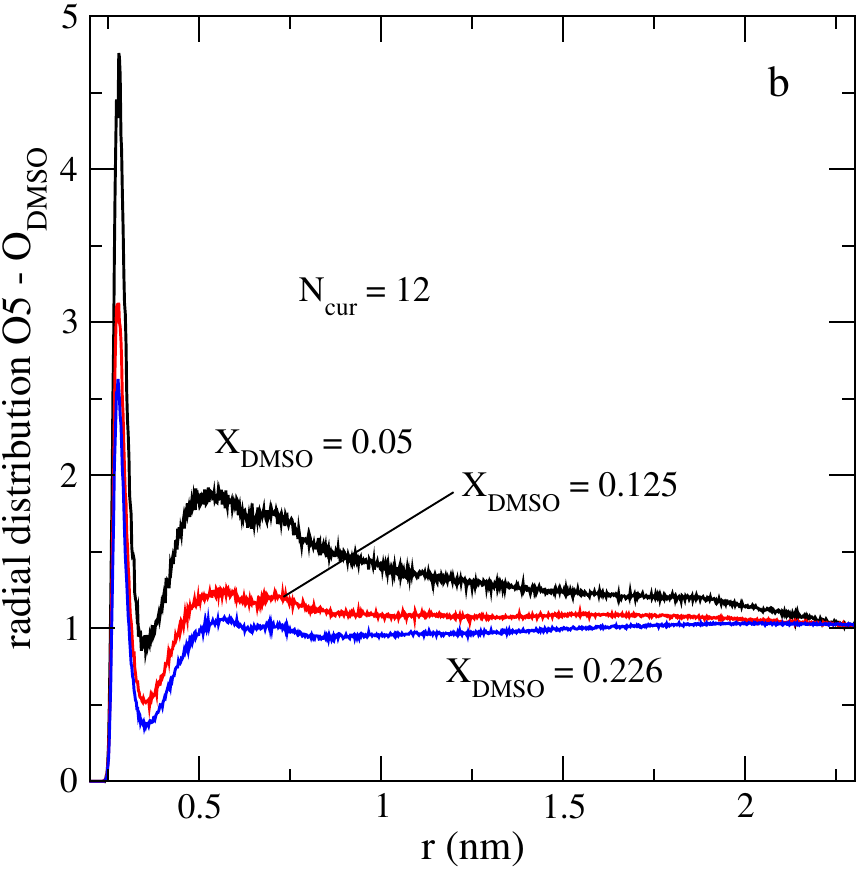}
\includegraphics[height=5.5cm,clip]{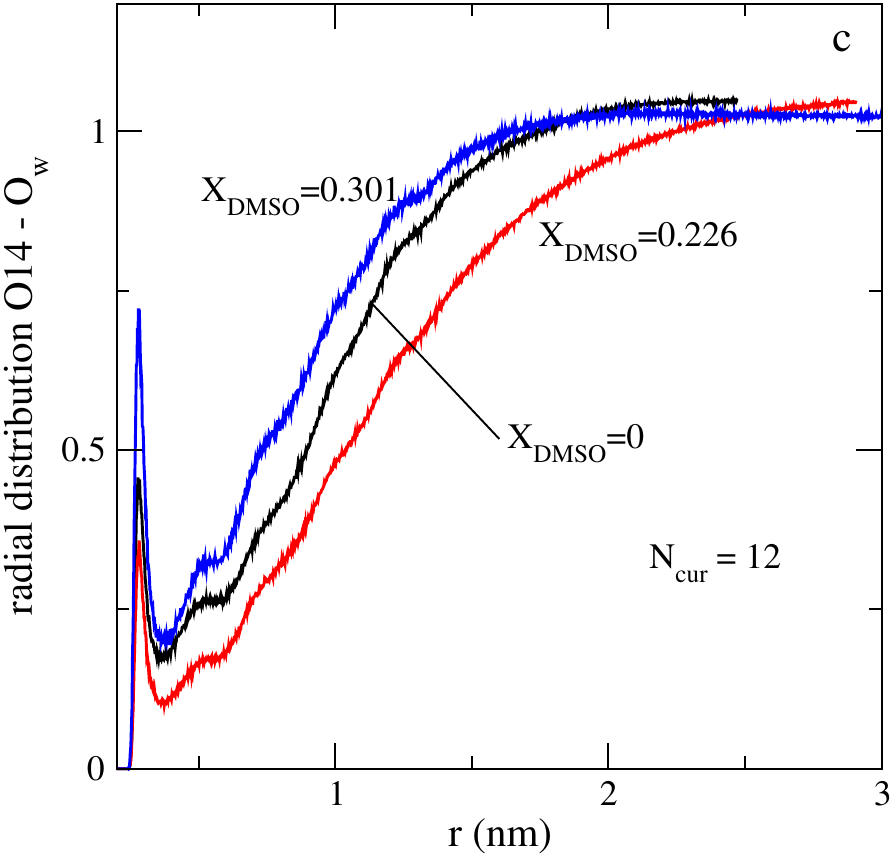}
\includegraphics[height=5.5cm,clip]{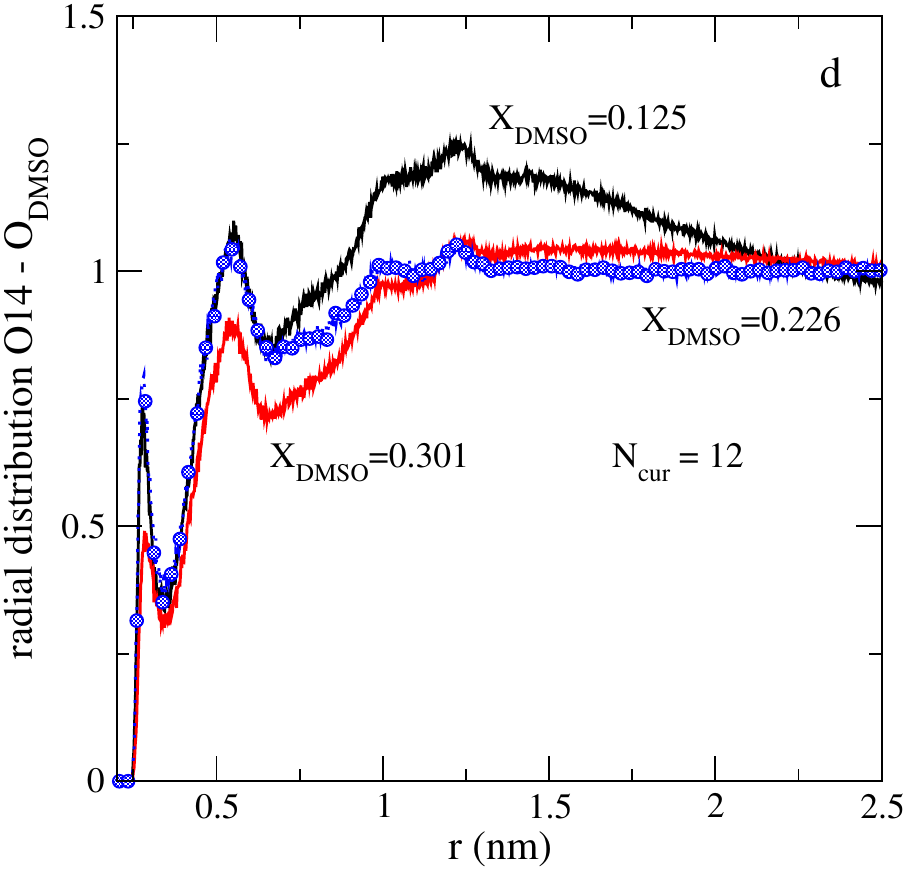}
\caption{(Colour online) Radial distribution of oxygen atom of water molecules
and of DMSO molecules, respectively, around O5 and O14 oxygen atoms of
curcumin molecules ($N_{\text{cur}}$ = $12$)
at different concentration of DMSO species in the solvent.
The values of $X_{\text{DMSO}}$ are given in both panels of the figure.
}
\label{fig12}
\end{figure}

disaggregation of curcumin clusters upon increasing the DMSO fraction
results in the changes of the curcumin molecules-solvent interface.
The water species at  $X_{\text{DMSO}} = 0$ prefer to avoid curcumin molecules 
as it follows from panels~a and c of figure~\ref{fig12}. Similar trends preserve, if
the fraction of DMSO increases to $X_{\text{DMSO}} = 0.226$. If a very small amount
of water is substituted by DMSO in the solvent (panel~b, $X_{\text{DMSO}} = 0.05$), all these DMSO
molecules permeate the curcumin cluster and/or surround the O5 exposed oxygen of 
curcumin molecule. This result seems to 
confirm a combined and local (because it
refers to O5 oxygen peripheric site) hydrophobic effect when both the curcumin and DMSO molecules
avoid contacts with water species. The O$_{\text{DMSO}}$--O$_{\text{DMSO}}$ distribution
is not shown for economy of space.
At a higher DMSO fraction, see the red and blue
lines in panel~b, the balance of interactions becomes different. There exists
competition between DMSO preference for curcumin and mixing trends of DMSO with water.
Consequently, some of the DMSO molecules ``dissolve'' in curcumin (see the first maximum
at $r \approx 0.2$~nm) whereas the others are almost uniformly distributed at larger
distances. Concerning the distribution of the DMSO oxygens with respect to 
the central fragment of a curcumin molecule (panel~d of figure~\ref{fig12}), we see that the DMSO
species fill the space around O14 in absence of O$_w$ (cf. panels~c and~d),
at a larger separation the distribution is rather uniform, unless the curcumin cluster
is present (like in the case at $X_{\text{DMSO}} = 0.125$ in panel~d).

\begin{figure}[!ht]
 \centering
\includegraphics[height=5.5cm,clip]{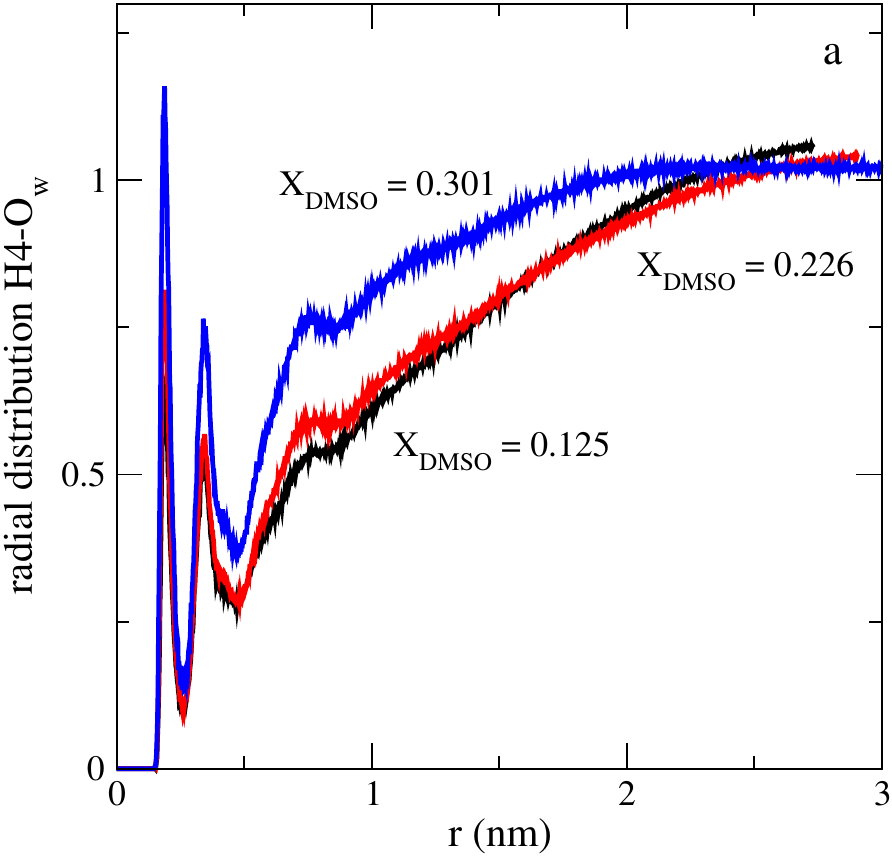}
\includegraphics[height=5.5cm,clip]{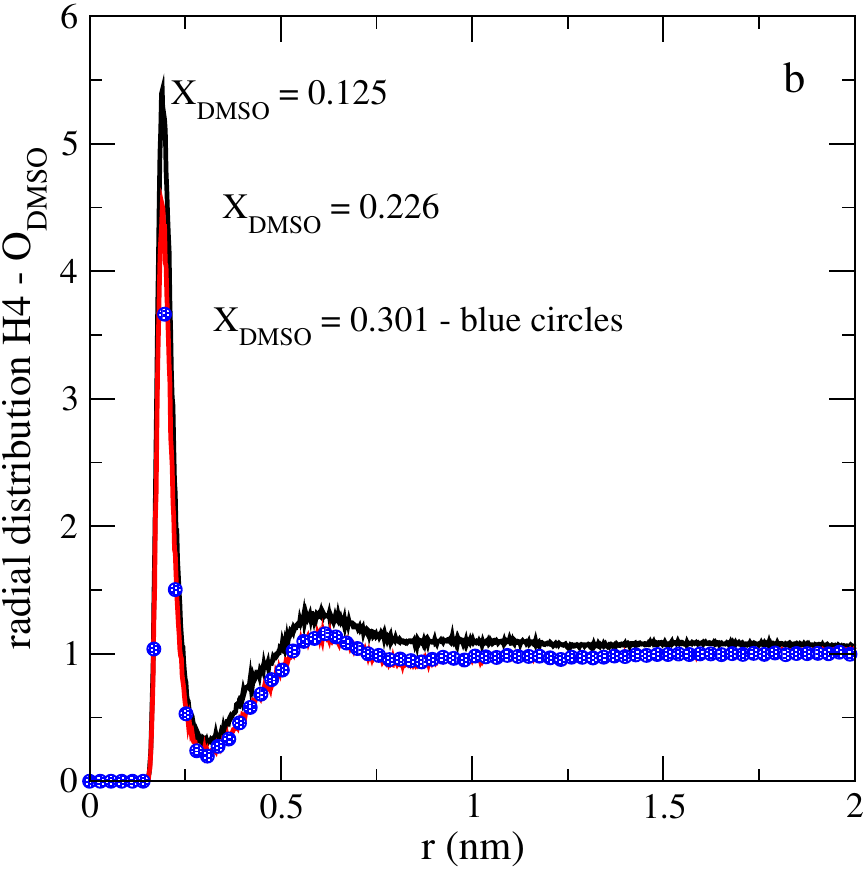}
\includegraphics[height=5.5cm,clip]{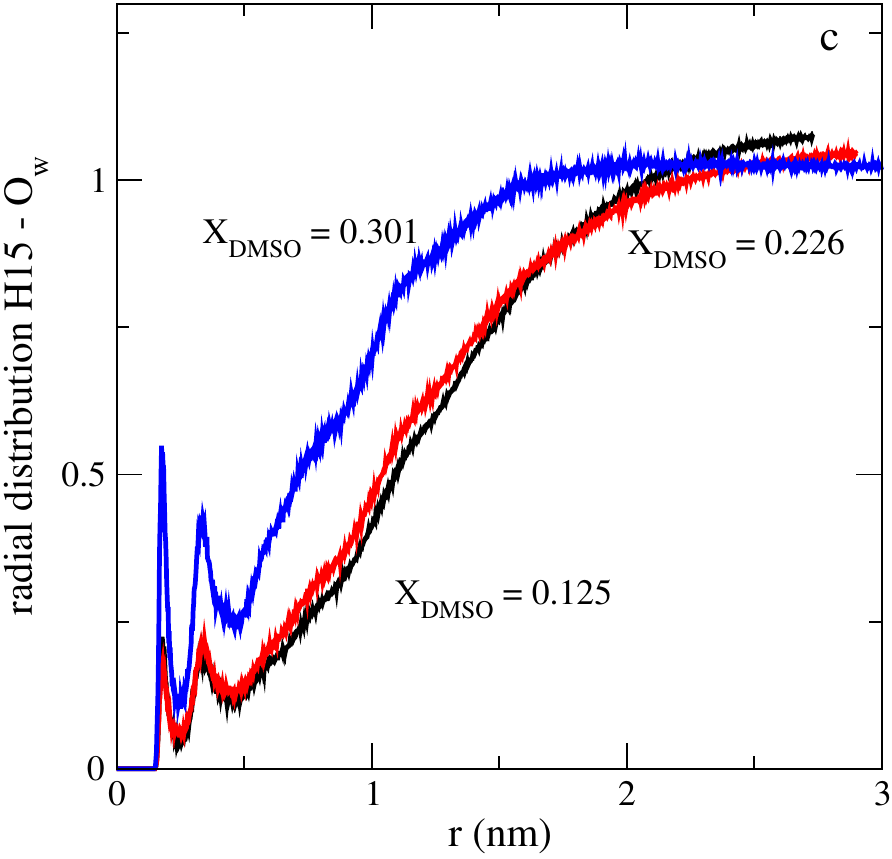}
\includegraphics[height=5.5cm,clip]{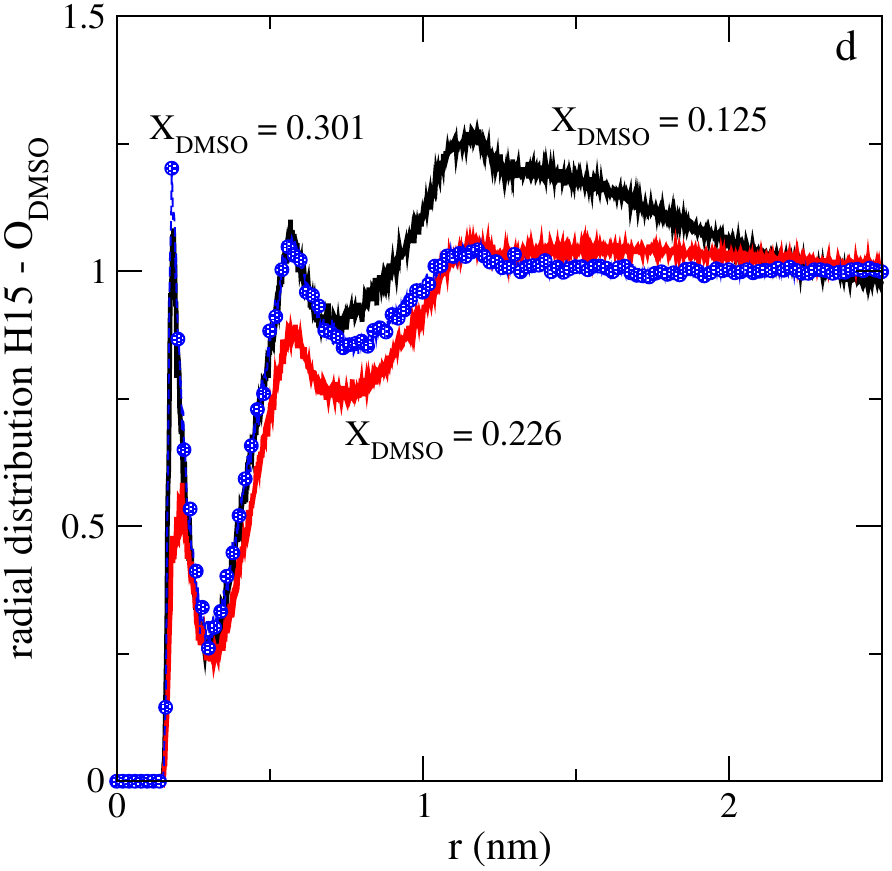}
\caption{(Colour online) Radial distribution of oxygen atom of water and of DMSO molecules
respectively, around H4 and H15 hydrogens of
curcumin molecules ($N_{\text{cur}}$ = $12$)
at different concentration of DMSO species in the solvent.
The values of $X_{\text{DMSO}}$ are given in both panels of the figure.
}
\label{fig13}
\end{figure}

In figure~\ref{fig13}, we present similar distributions of water, O$_w$, and DMSO, O$_{\text{DMSO}}$ oxygens,
although around H4 and H15 hydrogen atoms of a curcumin molecule. 
The probability of finding water oxygen close to H4 is low (panel~a). It  slightly increases
if the DMSO fraction in the solvent increases. By contrast, the probability for
H4-O$_{\text{DMSO}}$ contact is quite high, possibly even with the formation of hydrogen bond
(cf. panel~a and panel~b). From panel~c of this figure~\ref{fig13}, we learn that water
molecules can hardly approach H15 hydrogen of the curcumin. In fact, the depletion
region of this distribution becomes less pronounced upon increasing the DMSO
fraction in the solvent, i.e., when the curcumin clusters disaggregate. From panel~d
of the figure, one can conclude that trends for DMSO permeation into curcumin
clusters, namely to become close to H15, are weak. Rather, the DMSO species prefer
to locate on the surface of cluster entities (at $X_{\text{DMSO}} = 0.125$) or to
distribute uniformly at a distance above $r > 1.0$nm. In general, we observe that 
the distribution of both solvent species, water and DMSO, is rather heterogeneous
either over the curcumin cluster surface or in the solvation shell of curcumin
monomers.

\newpage
\subsection{On the relation with experimental observations}

It is known that curcumin is soluble in various solvents~\cite{lv,cui,Priyadarsini-2009}. 
In particular, in ethanol its solubility is of $10$~mg/ml. On the other hand,
its solubility on DMSO is more than twice higher, $25$~mg/ml. 
The CWAP for curcumin in water-ethanol solvent
is at volume/volume $50\%$, \cite{pereira}. 
However, these authors have performed analysis of 
the aggregation behaviour for the curcumin solution at a single dye
concentration. Moreover, it has been observed that
the tautomeric equilibrium may interfere into the aggregation of curcumin
species upon changing the solvent composition.  

\begin{figure}[!ht]
 \centering
\includegraphics[height=5.5cm,clip]{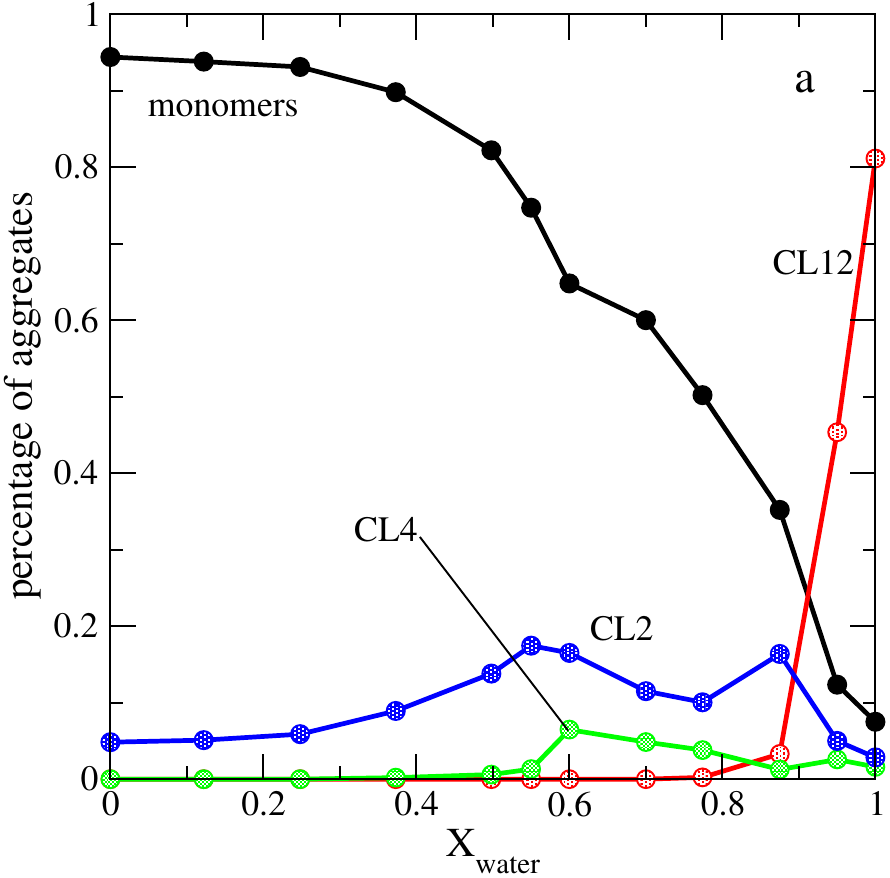}
\includegraphics[height=5.5cm,clip]{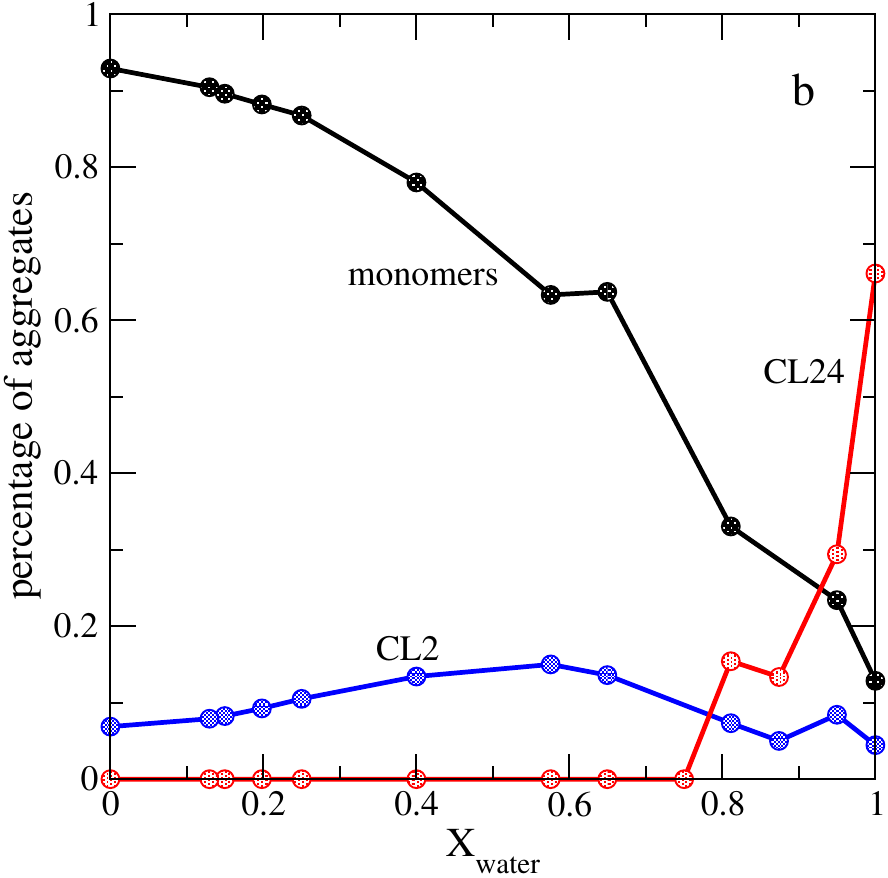}
\caption{(Colour online) 
Panel~a: Fractions of monomers and of clusters with $2$, $4$ and $12$ curcumin molecules
in the system with  $N_{\text{cur}} = 12$. 
Panel~b: Fractions of monomers and of clusters with $2$ and $24$ curcumin molecules
in the system with  $N_{\text{cur}} = 24$.
}
\label{fig14}
\end{figure}

We  normalized our data for the histogram distribution
of cluster sizes and converted the composition scale into the number fraction of water
rather than the DMSO.
Then, the results for the system with $12$ curcumin molecules look as shown
in panel~a of figure~\ref{fig14}. We observe that the fraction of curcumin monomers overcomes
the fraction of clusters with $12$ molecules at $\approx 0.9$ (from the 
crossing of red and black lines in each panel). This
fraction rapidly decays indicating that the smaller clusters are present in the
system. However, the fraction of even quite small clusters composed of $4$ 
curcumin molecules is very small and exhibits a discontinuous jump down to
almost zero, if water concentration decreases around $0.6$. Thus, the
the curcumin molecules are predominantly in their monomer form
in an ample interval, apparently at $X_{\text{water}} < 0.85$.
If a more concentrated solution of 
curcumin in water-DMSO solvent (twice higher) is considered, the fractions of 
monomers, dimers and of a cluster with $24$ curcumin molecules behave
as it is shown in panel~b of figure~\ref{fig14}. 
If we analyze the trends of behaviour of the coordination numbers, the
transition interval between monomers and clusters is expected to occur 
in the interval between  $0.7$ and 
$0.875$ of water contents, in terms of $X_{\text{water}}$. 
Actually, one should consider larger systems to establish 
more precisely, if the running coordination number saturates to the number
of curcumin molecules. 
Still another possibility is to establish this interval 
from the behaviour of the self-diffusion coefficient of curcumin species
in the solutions. If the cluster motif dominates, the diffusion is
expected to be very low. By contrast, at a higher DMSO contents (or quite low
water fraction), when the monomer fraction dominates, one would expect
quite reasonable diffusion of curcumin species.

Anyway, apparently there exists correlation between the solubility of curcumin 
in a solvent blend and the location of the CWAP.

\section{Summary and conclusions}

At this stage of the project, we applied a quite simple modelling of curcumin solutions in 
water-DMSO solvent of variable composition. Molecular dynamics computer simulations were
performed at room temperature and ambient pressure. Our principal findings follow from
the analyses of solely the microscopic structure of the system
while changing the solvent composition. 
More specifically, we focused on the disaggregation of curcumin
clusters upon substituting a certain amount of water molecules of the solvent by the DMSO species.
We described the conditions at which curcumin monomers dominate the structure of solution
rather than their clusters. Our finding follow from the COM-COM distribution of curcumin
species and a set of auxiliary descriptors.
However, a more refined picture for the mechanism
of degradation of clusters is necessary. Namely, it would be desirable to complement
the present investigation by the description of hydrogen bonding network 
between different species, its time evolution and the corresponding life times 
for different solvent compositions. 
Usually, as documented in the literature, geometric and energetic criteria for H-bond 
existence can be applied. These types of criteria formally can be developed or mapped to 
the level of clusters. This non-trivial task requires additional efforts, however. 
Moreover, we expect that the degradation of clusters can be described in terms of 
various dynamic properties, e.g., of different auto-correlation functions. Then, 
more comprehensive insights into the
relation between experimental spectroscopic observations and theoretical
predictions can be obtained. 
Description of thermodynamic repercussions, including changes of miscibility
and solvation phenomena, should be studied for systems in question as well.

A more sophisticated  modelling of the curcumin solutes
(in various aspects, e.g., by taking into account the tautomeric equilibrium, 
stability or instability of the molecule) and of solvent species is undoubtedly necessary.
Intuitively, similar trends of behaviour, as we observe in the present work,  should be
expected for curcumin solutions in water-simple alcohols solvent blends. Then,
additional insights into the dependence of behaviour of curcumin solutions 
on the nature of co-solvent would be reached. These issues are under study in our laboratory.

\section*{Acknowledgements}
O.P. is grateful to M. Aguilar for technical support of this
work at the Institute of Chemistry of UNAM. Fruitful discussions with Dr. Manuel Soriano
at the Institute of Chemistry of UNAM during the early stage of this project  
are gratefully acknowledged. T.P. acknowledges allocation of
computer time at the cluster of the Institute for Condensed Matter Physics of NAS of Ukraine and
the Ukrainian National Grid.

\newpage
\ukrainianpart

\title{Аспекти мікроскопічної структури куркуміну у водно-диметилсульфоксидному розчиннику. Дослідження комп’ютерним моделюванням методом молекулярної динаміки.}
\author{Т. Пацаган\refaddr{label1}\refaddr{label2}, О. Пізіо\refaddr{label3}}

\addresses{
	\addr{label1} Інститут фізики конденсованих систем Національної академії наук України\\
	79011, м. Львів, вул. Свєнціцького, 1, Україна
	\addr{label2} Інститут прикладної математики та фундаментальних наук, Національний університет ``Львівська політехніка'', Україна, 79013, Львів, вул.~С.~Бандери,  12
	\addr{label3} Інститут хімії, Національний автономний університет Мехіко, 
	Circuito Exterior, 04510, Мехіко, Мексика}

\makeukrtitle

\begin{abstract}
	Аспекти мікроскопічної структури куркуміну у водно-диметилсульфоксидному розчиннику. Дослідження комп’ютерним моделюванням методом молекулярної динаміки.
	
Ми досліджуємо деякі аспекти мікроскопічної структури куркуміну у водно-диметилсульфоксидному розчиннику різної композиції. Для цього проведено комп’ютерне моделювання методом молекулярної динаміки при постійній температурі та тиску. Система складається з OPLS-UA моделі для енольного конформера куркуміну (J. Mol. Liq., \textbf{223}, 707, 2016), OPLS моделі для диметилсульфоксиду (DMSO) і SPC/E моделі води. Розраховано радіальні розподіли центрів мас молекул куркуміну та проаналізовано відповідні до них біжучі координаційні числа. З'ясовано вплив збільшення концентрації DMSO у розчиннику вода-DMSO на дезагрегацію кластерів куркуміну. Досліджено зміни розподілу молекул води та DMSO навколо молекул куркуміну. Виконано якісне порівняння наших вислідів з результатами інших авторів. Обговорено можливість взаємозв'язку результатів моделі з експериментальними даними в термінах так званого критичного відсотка агрегації води.
	\keywords куркумiн, модель об'єднаних атомiв, молекулярна динамiка, вода, диметилсульфоксид, кластери
\end{abstract}

\lastpage
\end{document}